\documentclass{moriond}

\usepackage{amssymb}

\def\be{\begin{equation}}
\def\ee{\end{equation}}
\def\bea{\begin{eqnarray}}
\def\eea{\end{eqnarray}}

\newcommand{\hc}{\mathrm{h.c.}}
\newcommand{\sun}{\mathrm{SU}(N)_{\mathrm{DC}}}

\newcommand{\ldc}{\Lambda_{\mathrm{DC}}}
\newcommand{\ndc}{N_{\mathrm{DC}}}
\newcommand{\mtcb}{m_{\mathrm{DCb}}}

\newcommand{\udb}{\mathrm{U}(1)_{\mathrm{DB}}}
\newcommand{\tcp}{DC$\pi~$}

\newcommand{\Photo}{}

\begin{document}
\vspace*{4cm}
\title{ASYMMETRIC ACCIDENTAL COMPOSITE DARK MATTER}

\author{SALVATORE BOTTARO, \textbf{MARCO COSTA}}
\address{Scuola Normale Superiore, Piazza dei Cavalieri 7, 56126 Pisa, Italy,
\\INFN Sezione di Pisa, Largo B. Pontecorvo 3, I-56127 Pisa, Italy}
\author{OLEG POPOV}
\address{Department of Physics, Korea Advanced Institute of Science and Technology, 291 Daehak-ro, Yuseong-gu, Daejeon 34141, Republic of Korea}

\maketitle
\abstracts{The goal of this work is to find the simplest UV completions of Accidental Composite Dark Matter models that can dynamically generate an asymmetry for the Dark Matter candidate, the lightest \emph{dark baryon}, and simultaneously annihilate the symmetric component. In this framework Dark Matter is an accidentally stable bound state of a confining $\sun$ gauge group that can interact weakly with the visible sector.
The generation of asymmetry for such candidate happens via the out-of-equilibrium decay of two flavors of a heavy scalar $\phi$, with mass $M_\phi > 10^{10}$ GeV. Below such scale, the models recover accidental stability, or long-livedness, of the Dark Matter candidate. The symmetric component is annihilated by residual confined interactions provided that the mass of the dark baryon $\mtcb < 75$ TeV.} 
\section{Introduction}
The nature of Dark Matter (DM) is unknown, and its understanding is one of the major problems of fundamental physics. 
Current experiments indicates that its abundance $\Omega_{\mathrm{DM}}$ is close to the baryonic one $\left(\Omega_b\right)$ \cite{Aghanim:2018eyx}:
\begin{equation}\label{eq:dm_abundance}
\Omega_{\mathrm{DM}}h^2=0.11933\pm0.00091s\simeq 5 \Omega_b h^2 \,.
\end{equation} 
Despite the huge experimental effort, little is known about DM properties, except that it must be stable (or very long-lived at least), and interacting weakly with Standard Model (SM) particles.\\
Stability of DM can be explained by replicating the success of the SM in predicting proton stability. If the DM is a composite particle charged under a new gauge interaction, gauge invariance can lead to the existence of an accidental symmetry that stabilizes the DM, like the \emph{dark baryon number} $\udb$.
This suggests to explore models of Composite DM, in which the DM is a composite object stabilized by an \emph{accidental symmetry} of the low energy theory.\\
In recent years, Composite DM models have been studied in their asymmetric version \cite{Petraki:2013wwa,Zurek_2014} (ADM). In ADM models only the DM particles make up the observed abundance, while their antiparticles are absent.
This scenario has a distinct phenomenology with respect to the symmetric case. Indeed the absence of residual annihilations and the presence of extra intractions in the Dark Sector (DS) can lead to the formation of bound states (BSF) of DM particles like \emph{dark nuclei}, both in the early \cite{Redi:2018muu,Krnjaic:2014xza} and late universe. BSF has peculiar Indirect Detection signatures if the DS has electroweak interactions under the SM \cite{Mahbubani:2020knq,Mahbubani:2019pij}.
ADM models can also shed light on the coincidence between the visible and DM energy density of Eq.\ref{eq:dm_abundance}, by relating the baryon and DM asymmetry via some mechanism, like the decay of a heavy particle.\\
However, as pointed out by Sakharov \cite{Sakharov:1967dj}, in order to generate an asymmetry in the DS the $\udb$ responsible for stabilizing the DM must be broken. This can potentially make the DM decay too fast, especially for  multi-TeV masses. The goal of this work \cite{Bottaro:2021aal} is to explore how the ADM paradigm can be reconciled with the accidental stability found in Composite DM models.
\section{Accidental Composite Dark Matter models}
Our starting points are Accidental Composite Dark Matter models \cite{Antipin:2015xia}.
These models feature new vector-like fermions, the \emph{dark quarks} $\Psi$ (DCq), with mass $m_\Psi$ and charged in the fundamental of a new QCD-like confining gauge group $\sun$, the \emph{dark color} (DC).
Below the DC confinement scale $\ldc\gg m_\Psi$, the DCq's bind into stable \emph{dark baryons} (DCb) and unstable light \emph{dark pions} (\tcp).
The DCb stability is ensured by the accidental \emph{dark baryon number} $\udb$ under which each DCq is charged by one unit.\\
Yukawa interactions between DCq's and the Higgs break the species number of the various DCq and allow the decay of \tcp 's.
Notice that in general the DCq's will be charged under the SM. The only requirement is that their SM charges are such that the DM particle (\emph{i.e.} the lightest stable DCb) is neutral, or at most weakly charged.
This implies that light, unstable, SM-charged \tcp 's can be  produced at colliders. This sets a lower bound on the DCb mass $\mtcb > \mathcal{O}(1)$ TeV, except for the model containing only the SM singlet DCq $N$.\\
The mass of the DCb is set by matching the observed relic abundance.
The DCb is kept in chemical equilibrium with its antiparticle ($\overline{\mathrm{DCb}}$) via non-perturbative annihilations.
This cross section can be parametrized as $\langle\sigma v\rangle\approx 100/\mtcb^2$.
The observed $\Omega_{\mathrm{DM}}$ is obtained for $\mtcb \approx 100$ TeV.

\section{Asymmetric Accidental Composite Dark Matter models}
\subsection{Asymmetric abundance}
Consider now a scenario with a primordial asymmetry of DCb
$\eta_{\mathrm{DM}}\equiv(n_{\mathrm{DCb}}-n_{\overline{\mathrm{DCb}}})/s$,
where $n_{\mathrm{DCb}},n_{\overline{\mathrm{DCb}}}$ are the number densities of the DCb and its antiparticle respectively, and $s$ is the entropy density.
The DM abundance is set only by the DCb's provided $n_{\overline{\mathrm{DCb}}}\ll n_{\mathrm{DCb}}$:
\begin{equation}\label{eq:asym_abundance}
\Omega_{\mathrm{DM}}\propto\eta_{\mathrm{DM}} \mtcb \;. 
\end{equation} 
In this scenario the annihilation cross section must be larger than in the symmetric case in order to annihilate the residual symmetric component. While in other models in the literature this condition requires extra interactions, in the framework of ACDM the non-perturbative annihilations can provide such large cross sections. By properly solving the Boltzmann equations for the symmetric and asymmetric components \cite{Graesser:2011wi}, it is possible to show that if $\mtcb < 75$ TeV, the former drops  below 1\% of the total observed density, making the model asymmetric.
\subsection{Generating the asymmetry}
\begin{figure}[!ht]
\begin{minipage}{0.32\linewidth}
\centerline{\includegraphics[width=0.7\linewidth]{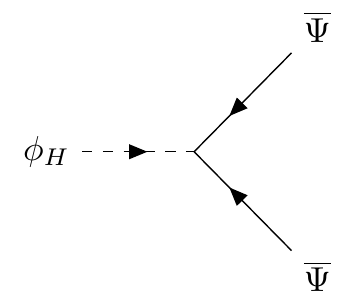}}
\end{minipage}
\hfill
\begin{minipage}{0.33\linewidth}
\centerline{\includegraphics[width=0.8\linewidth]{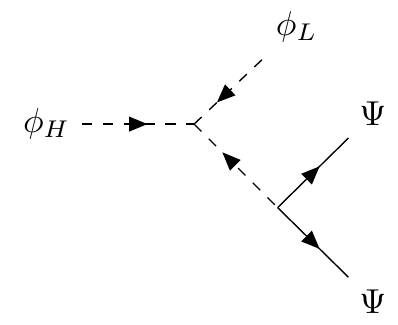}}
\end{minipage}
\hfill
\begin{minipage}{0.33\linewidth}
\centerline{\includegraphics[width=0.85\linewidth]{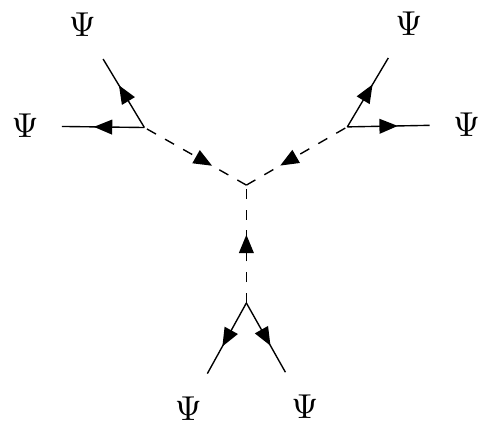}}
\end{minipage}
\caption[]{Examples of $\udb$-violating processes involving the scalar $\phi$ in the $\phi \Psi \Psi + V$ models. Arrows denote $\udb$-number flow. Dashed (solid) lines indicate scalars (fermions). \textbf{Left and Center}: decays of the scalars into the DCq's, responsible for generating $\eta_{\mathrm{DM}}$. \textbf{Right}: process responsible for the oscillation-inducing Majorana mass of the DCb ($\Psi^3$ in this case).}
\label{fig:decays}
\end{figure}

The models still lack a way to generate the asymmetry $\eta_{\mathrm{DM}}$.
Any UV completion aiming at doing so must break the stabilizing $\udb$.
In the UV theory, restricting to renormalizable terms, no $\udb$-breaking operator can be written by adding only fermions. This forces to add a new scalar $\phi$ charged under $\sun$.
The possible scalar interactions can be classified in three categories:
\begin{equation}\label{eq:phi_terms}
\phi \Psi \psi_{\mathrm{SM}}\;, \quad \phi \Psi \Psi\;, \quad V(\phi,H)\;,
\end{equation}
where $\psi_{\mathrm{SM}}$ is a generic SM fermion, and in the second Yukawa the two DCq's are contracted so that the bilinear carries a net $\udb$ charge. $V$ is a potential term that contains interactions violating $\phi$-number:
\begin{equation}\label{eq:v_phi}
V(\phi,H) \;\supset \quad
\phi^3\;, \quad  \phi^4\;, \quad \phi^3H^* \;.
\end{equation}
If only a single term of the one in Eq.\ref{eq:phi_terms} is present, it is always possible to consistently assign a $\udb$ charge to $\phi$ such that $\udb$ is conserved, thus preventing the production of a net asymmetry. Therefore the presence of at least two of such operators is required. From this, together with the assumption that the DCq's are in the fundamental of $\sun$, it follows that $\ndc=3,4$. 
Out of all the possibilities, we found two viable classes of models:
\begin{itemize}
\item $\phi \Psi \Psi + V(\phi)$. The interactions require $\phi$ to be in the $2$-symmetric representation of $\sun$. The models predict an exactly stable DCb via an accidental $\bf{Z}_2$ symmetry under which the DCb is charged. There is no link with the baryonic asymmetry $\eta_b$. The DCb's can oscillate into their antiparticles, provided $M_\phi<10^{13}$ GeV.
\item $\phi \Psi \psi_{\mathrm{SM}} + V(\phi)$. The scalar is in the (anti)fundamental of $\sun$. The models predict unstable, but very long-lived DCb's. Since in the model $\eta_b\sim\eta_{\Psi}$, a simultaneous explanation of the two abundances is possible only for $\mtcb\simeq 5$ GeV. This is allowed by collider bounds only if the DCq is the SM singlet $N$. By properly charging $\phi$ under the SM it is possible to obtain the needed interactions for this case:
\begin{equation}
\lambda \phi^3 H^*+y\phi^\dagger q_L N\; + \hc \quad \left(\ndc=3 \right)\;\;.
\end{equation}
\end{itemize}
Models realizing the breaking of $\udb$ via two Yukawas lead to DCb decay faster than the bounds on DM lifetime  \cite{Cohen:2016uyg} $\tau^{-1}\le 10^{-53}$ GeV, provided $\mtcb\simeq \mathcal{O}(10)$ TeV, $\mathcal{O}(1)$ couplings and $M_\phi$ lower than the Planck scale.\\
The next step is to show that the interactions introduced are enough to generate the proper amount of asymmetry $\eta_{\mathrm{DM}}$ needed to satisfy Eq.\ref{eq:asym_abundance}. By introducing two flavors of the scalar $\phi_{H,L}$ with similar masses $M_{\phi_H}\gtrsim M_{\phi_L}=M_\phi$, there is enough CP violation to guarantee that the out-of-equilibrium decays of the $\phi$'s, shown in Fig.\ref{fig:decays} (left and center), will produce enough asymmetry for masses in the  $\mtcb < 75$ TeV range. The wash-out of the asymmetric component can be avoided for technically natural values of the couplings starting from $M_\phi>10^{10}$ GeV.

\section{Phenomenology}
As remarked before, Composite Asymmetric Dark Matter models have peculiar phenomenological signatures that distinguish them from their symmetric counterpart: BSF \cite{Mahbubani:2019pij} leads to specific $\gamma$-lines, and $\mtcb$ is lighter in the ADM case.
Compositeness in principle can be probed via gravitational waves coming from the confinement phase transition \cite{Schwaller:2015tja}. Their typical frequencies will be different with respect to the symmetric case due to the smaller $\ldc$.
Compositeness also implies that the DM has magnetic (and possibly electric) dipoles that can be tested at Direct Detection experiments such as XENON.\\
The asymmetry generation mechanism \emph{i.e.} the scalar $\phi$, can be probed in the $\phi\Psi \Psi + V$ model. Indeed the low energy theory includes a $\udb$-violating DCb Majorana mass, shown in Fig.\ref{fig:decays} (right). If $M_\phi<10^{13}$ GeV, this term can induce oscillations between the DCb and its antiparticle, and lead to detectable late time residual annihilations. This process can in principle wash-out the asymmetry of free DCb's. The fraction of DCb's that are bound in dark nuclei will not oscillate: the binding energies inside a dark nucleus are different between the DCb and $\overline{\mathrm{DCb}}$.

\section{Conclusions}
In this work we built UV completions of Accidental Composite Dark Matter models that lead to an Asymmetric Dark Matter scenario.
In order to do so, the stabilizing $\udb$ must be broken. 
This is done by introducing a heavy scalar, whose decay generates the correct DM asymmetry. The symmetric component is annihilated by residual confined interactions between the DCb's and their antiparticles. We found two possible classes of models. The first predicts a stable DCb, that can oscillate into its antiparticle (provided it is not bound in dark nuclei). The second class predicts unstable but long-lived DM. A simultaneous explanation of $\Omega_b$ implies $\mtcb$ in the GeV range, which is allowed only in the $\Psi=N$ model, with suitable SM charges for $\phi$. We briefly discussed the phenomenology of Asymmetric ACDM models, pointing out that they can be distinguished from other DM scenarios.

\section*{Acknowledgments}

The work of M.C. is partially supported by the PRIN 2017L5W2PT.

\section*{References}

\bibliography{marcocosta_bib}

\end{document}